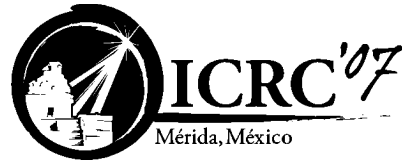

# Education and Outreach for the Pierre Auger Observatory

G. SNOW[1] FOR THE PIERRE AUGER COLLABORATION[2]
[1]*Department of Physics and Astronomy, University of Nebraska, Lincoln, NE 68588-0111 USA*
[2]*Av. San Martin Norte 304 (5613) Malargüe, Prov. de Mendoza, Argentina*
gsnow@unlhep.unl.edu

**Abstract:** The scale and scope of the physics studied at the Auger Observatory offer significant opportunities for original outreach work. Education, outreach, and public relations of the Auger collaboration are coordinated in a task of its own whose goals are to encourage and support a wide range of efforts that link schools and the public with the Auger scientists and the science of cosmic rays, particle physics, and associated technologies. This report focuses on the impact of the collaboration in Mendoza Province, Argentina, as: the Auger Visitor Center in Malargüe that has hosted over 29,000 visitors since 2001, the Auger Celebration and a collaboration-sponsored science fair held on the Observatory campus in November 2005, the opening of the James Cronin School in Malargüe in November 2006, public lectures, school visits, and courses for science teachers. As the collaboration prepares the proposal for the northern Auger site foreseen to be in southeast Colorado, plans for a comprehensive outreach program are being developed in parallel, as described here.

## Introduction

Education and public outreach have been an integral part of the Auger Observatory since its inception. The collaboration's outreach activities are organized in a dedicated Education and Outreach Task that was established in 1997. This Task has been particularly active since major construction and deployment activities began for the southern hemisphere site on the Pampa Amarilla in Mendoza Province, Argentina. With the Observatory headquarters located in the remote city of Malargüe, population 20,000, early outreach activities included public talks, visits to schools, and courses for science teachers and students. These were aimed at familiarizing the local population with the science of the Observatory and the presence of the large collaboration of international scientists in the isolated communities and countryside of Mendoza Province. With the assistance of the Malargüe Office of Tourism and municipal authorities, the collaboration has been successful becoming part of the local culture. As an example of the Observatory's integration into local traditions, the collaboration has participated in the annual Malargüe Day parade since 2001 with collaborators walking behind a large Auger banner. The Observatory's outreach efforts have been documented in previous ICRC proceedings contributions [1]. We report here highlights of recent education, outreach, and public relations efforts.

## The Auger Visitor Center in Malargüe

The Auger Visitor Center (VC), located in the Observatory's office complex and data-acquisition facility and described in detail in [1], has continued to be a popular attraction for tourists and school groups since its opening in 2001. Through the end of April 2007, the VC has hosted over 29,000 visitors. The majority of visitors are from Argentina (~95%) and the remainder come from over 25 countries worldwide. Fig. 1 shows the cumulative number of visitors logged per year from November 2001 through April 2007.

Recent additions to the VC include (i) a scale model of the Observatory donated by the Forschungszentrum Karlsruhe that displays recorded air shower events in real time using computer-controlled LEDs at each surface detector and fluorescence detector location and (ii) a large flat-screen monitor and PC that allow visitors to tour all the Observatory's structures from above using customized Google Earth tools developed by Auger collaborators [2]. When "flying over" any detector or building in the latter display, an explanatory pop-up box is available in English and Spanish. An example of a screenshot obtained with Google Earth is shown in Fig. 2. A visitor can also choose to superpose 3D representations of real high-energy cosmic-ray events detected with the Observatory on the Google Earth display.

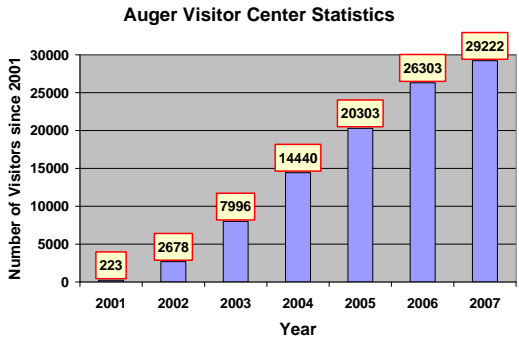

Figure 1: Cumulative number of people logged at the Auger Visitor Center from November 2001 through April 2007.

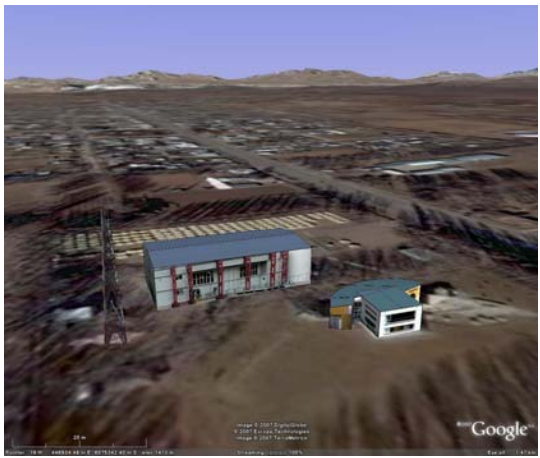

Figure 2: Google Earth image of the Auger office complex and assembly building in Malargüe.

## The Auger Celebration

The collaboration hosted a major Celebration on Nov. 9-11, 2005, to mark the progress of the Observatory and the presentation of the first physics results at the ICRC 2005. Over 175 visitors from almost all of the collaborating countries traveled to Malargüe to attend the Celebration. The visitors (see Fig. 3) included administrators from collaborating institutions, representatives from funding agencies that have supported the Observatory, representatives from the Argentine embassies of collaborating countries, local and provincial government authorities, plus press and media teams. The 3-day event featured talks on the history and status of the Observatory, the unveiling of monument listing the collaborating countries, traditional folk music and dance performances, and opportunities for visitors to tour the vast Auger site. The Celebration was covered extensively in newspaper, magazine, and television pieces that appeared in several of the collaborating countries, including an article in the CERN Courier [3].

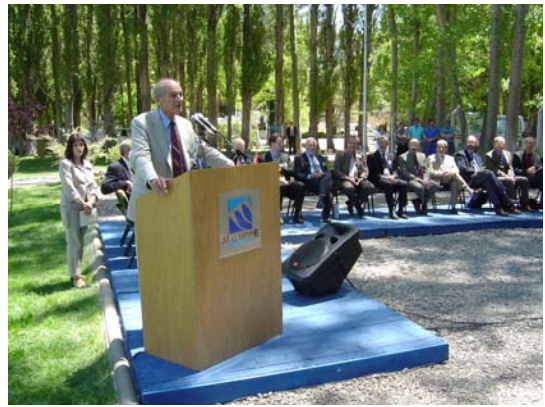

Figure 3: Emeritus spokesperson James Cronin addressing the assembled visitors and dignitaries at the Auger Celebration, November 2005.

## The Auger Science Fair

The collaboration sponsored a Science Fair for teachers and students from high schools in Mendoza Province that was held in the Observatory's assembly building (see Fig. 4) on Nov. 11-12, 2005. Twenty nine student-teacher teams displayed research projects in physics, chemistry, or technology at the Fair, some teams traveling long distances to reach Malargüe. A team of Auger collaborators judged the projects on the basis of science content, oral and visual presentation, and the written report that accompanied each project.

The projects were judged to be quite sophisticated, and the top projects were awarded prizes. During the Science Fair, participants were treated to presentations about the Observatory given in the Visitor Center. The collaboration is indebted to the Observatory staff, the local organizing team of 4 science teachers, and the city of Malargüe for helping to make the Science Fair a success. A second Auger-sponsored Science Fair is planned for November 2007.

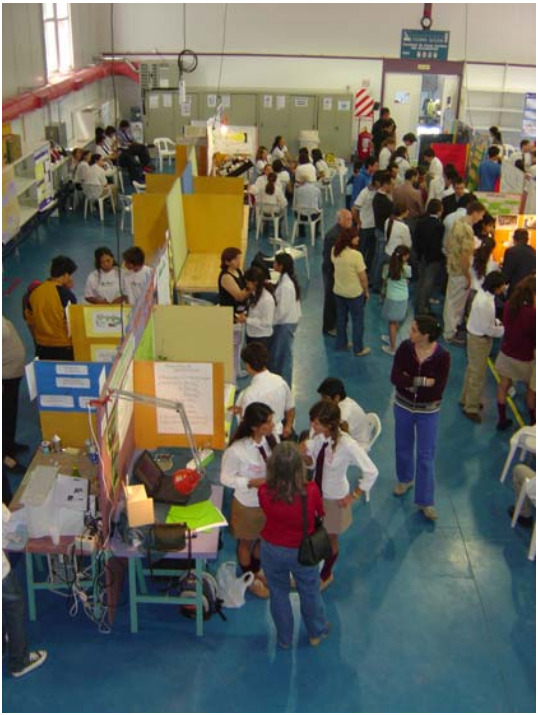

Figure 4: Participants preparing their displays at the Auger Science Fair, November 2005.

## Other Education/Outreach Activities

Members of the Outreach Task continue to offer informative talks and courses on inquiry-based science teaching techniques to students, teachers, and the general public. A Mexican collaborator, Rebeca López, is particularly active and is shown in Fig. 5 presenting the Observatory to the student body of a large secondary school in Malargüe. Another highlight was an excellent talk about cosmic rays and the Observatory given to English speaking students and teachers by Hans Blümer of Karlsruhe during the November 2006 collaboration meeting.

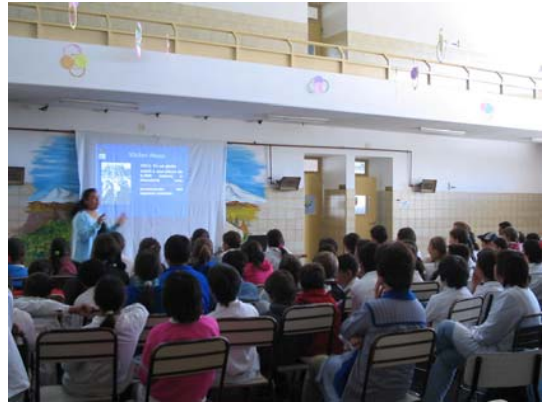

Figure 5: Rebeca López explaining the Observatory to students in Malargüe, November 2005.

The scholarship program which brings top Malargüe students to Michigan Technical University (MTU), described in [1], has enjoyed continued success. A $4^{th}$ student in the program will begin his studies in science or engineering in the fall 2007. The first MTU student from Malargüe who enrolled in 2001 has now completed a Masters Degree in mechanical engineering and has embarked on an engineering career in the U.S.

The collaboration recently decided to release 1% of the data collected by its surface array for outreach purposes. This data can be browsed on three dedicated web sites mirrored in Argentina, France, and the U.S. [4]. A complete reconstruction of the events is available, along with information from individual detector stations. Teaching modules to accompany the publicly released events are under development.

## The James Cronin School

A long-standing collaboration between the Observatory and the Province of Mendoza culminated in the building of a new secondary school in Malargüe that was inaugurated on November 16, 2006. The modern new building, shown in Fig. 6, was made possible, in part, by a generous donation from the Grainger Foundation in the U.S. that was secured by emeritus spokesperson James Cronin. The inauguration included a flag raising ceremony, followed by a number of speeches, including Cronin, Malargüe mayor Raúl Rodriguez, and Mendoza governor Julio Cobos. The event received wide press coverage. In the evening of the same day, the students and teachers from the school held a separate inaugural celebration that included student projects on display in the classrooms, plus dance and theatrical performances by the students.

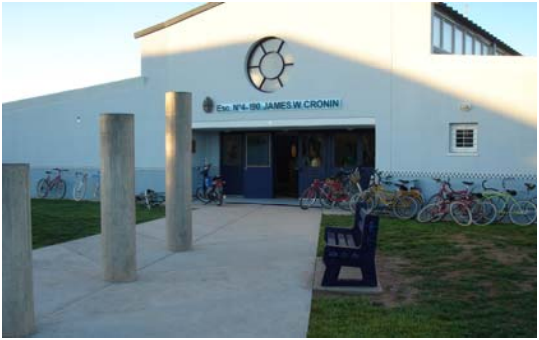
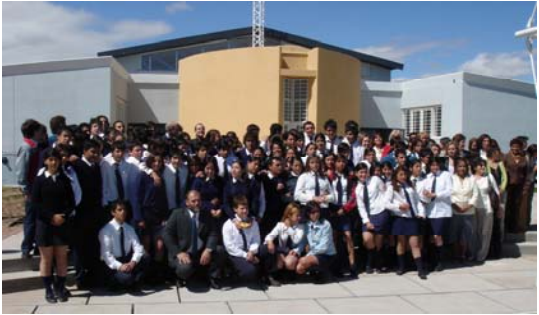

Figure 6: (Top) The front of the James Cronin School. (Bottom) Students and teachers behind the school on inauguration day.

## Northern Site Outreach

As the Auger collaboration prepares its proposal to build a northern hemisphere site in southeast Colorado, the Education and Outreach Task will lay the groundwork for a comprehensive outreach program that will be linked to the outreach efforts in the southern hemisphere. R&D for the northern site is foreseen to extend through 2008. Primary outreach goals in the R&D period are to promote the Observatory in southeast Colorado, provide information to people at all levels, and establish early ties with science teachers and students in the regional schools. Northern site outreach efforts began in earnest when Lamar Community College (LCC) hired Bradley Thompson, previously an international high school physics teacher, in January 2007 to spearhead the Colorado public outreach efforts in collaboration with the outreach Task Leader and other Auger members.

There will be three components to the outreach program during the northern site R&D phase. The first involves the public display of two surface detector stations and the opening of an interim VC at LCC, the anticipated headquarters of the proposed northern site. A public ceremony to unveil the surface detector stations and open the interim VC is planned for late July 2007. The second component will be a series of presentations given to regional schools and civic groups. These presentations will provide information about the Observatory to teachers, students, and civic leaders and will focus on the scientific, educational, and economic opportunities the Observatory may bring to the region. The third component targets high school students and teachers interested in cosmic ray physics, astrophysics, high-energy physics, and associated technologies. A small scale research program, modeled after the University of Nebraska's Cosmic Ray Observatory Project (CROP) [5], will be introduced at a small number of regional schools. A teacher professional development program is also scheduled to begin in the fall of 2007. This program will train teachers for the above research program, assess science education needs in the region, and probe opportunities to integrate cosmic ray and high-energy physics into the classrooms.